\documentclass{ws-p8-50x6-00}

\begin{document}

\title{Radial Correlations between two quarks}

\author{UKQCD Collaboration}

\author{A.M. Green, J. Koponen, P. Pennanen}

\address{Department of Physics and Helsinki Institute of Physics\\
P.O. Box 64, FIN--00014 University of Helsinki,Finland\\
E-mails: anthony.green@helsinki.fi,
jmkopone@rock.helsinki.fi, petrus@hip.fi}

\author{C. Michael}

\address{Department of Mathematical Sciences, University of Liverpool,
 L69 3BX, UK \\
E-mail: cmi@liv.ac.uk}


\maketitle

\abstracts{
In nuclear {\em many}-body problems the short-range correlation between
two nucleons
is  well described by the corresponding correlation in the {\em two}-body
problem. Therefore, as a first step in any attempt at an analogous
description of {\em many-quark} systems, it is necessary to know the {\em two-quark}
correlation. With this in mind, we  study the
light quark  distribution in a heavy-light meson with a static heavy quark.
The charge and matter radial
distributions of these heavy-light mesons are measured on a
 lattice  with a light quark mass about that of the strange quark.
Both distributions can be well fitted upto $r\approx 0.7$ fm with the
exponential form $w_i^2(r)$, where
 $w_i(r)=A\exp(-r/r_i)$.  For the charge(c) and matter(m) distributions
$r_c \approx 0.32(2)$fm and  $r_m \approx 0.24(2)$fm.  We also discuss
the normalisation of the total charge (defined to be unity in the
continuum limit) and matter integrated over all
space, finding 1.30(5) and 0.4(1) respectively for a lattice spacing
$\approx 0.17$~fm.
}

\section{Correlations between two nucleons}
\label{intro}
In nuclear many-body problems it is well known that the short-range
correlations between pairs of nucleons are often well described
by the corresponding correlations of the two-body problem. For example:
\begin{itemize}
\item {\bf Few nucleon systems.}
The NN correlations for the deuteron, $^3He$ and $^4He$, when
normalised to each other at their maxima, are essentially
indistinguishable upto about 2 fm. Only for larger internucleon
distances do the binding energy differences play a role \cite{glo}.
\item {\bf Many-nucleon systems.} Often variational methods are used
and a typical trial wavefunction is taken to be of the form
\be
|\Psi_T>= \left[\Pi_{i<j} F_{ij}\right] |\Psi({\rm Shell \ \ Model})>,
\ee
where $\Psi({\rm Shell \ \ Model)}$ is simply a product(antisymmetrised) of
single nucleon wavefunctions -- encoding the long range correlations.
However, the short range physics between the NN pairs is in the
\be
F_{ij}=\left[1+\sum_m u_m(r_{ij})O^m_{ij}\right],
\ee
where  $O^m_{ij}=[1,\sigma_i.\sigma_j,...]\bigotimes [1,\tau_i.\tau_j]$
are the various spin and isospin operators for two nucleons.
The variation then yields radial correlations
 $u_m(r_{ij})$ that are $\approx $ 2-body problem correlations -- see
for example Ref.~\cite{carl}.
\item {\bf Nuclear Matter} An extreme use in many-body systems of the
two-body correlation is the Moszkowski-Scott separation
method\cite{mos} in nuclear matter.
There the many-body wavefunction is taken to be of the form
\be
\begin{array}{lll}
\Psi({\rm Many-Body})&=\Psi({\rm Two-Body})& r< r_c \\
                     &=\sin (kr)             & r\ge r_c,\\
\end{array}
\ee
where $r_c$ is the value of $r$ at which $\Psi({\rm Two-Body})$ and
$\sin(kr)$ and their derivatives equal each other.
\end{itemize}
{\bf The conclusion is that, in many-nucleon systems, the two-nucleon
correlation from the two-body problem plays an important role. }

\vskip 0.2cm

The question can then be asked whether similar effects arise in
multi-{\em quark} systems. This may help in the understanding of
such systems. At present and in the forseeable future, since lattice
QCD calculations are  restricted
to very few quarks (upto four), some model {\em that encodes the
results of these few-quark calculations} is needed in order to proceed
to systems containing more quarks.

Earlier attempts by the present authors  have concentrated
on understanding the {\em energies} of four-quark systems in terms
of interquark potentials. These involved four heavy quarks
($Q^2\bar{Q}^2$)\cite{PG} and the interaction of two heavy-light mesons
($Q^2\bar{q}^2$) \cite{GKP}. In neither case  was it possible
 to understand the four-quark energies in terms of simply two-quark
potentials -- with the  result indicating that a four-quark form-factor was
necessary. This conclusion has also been supported by colour field 
measurements~\cite{PGM}. Here, as a first step, the emphasis is on extracting the {\em radial
correlation} between the two quarks of a heavy-light meson. Hopefully, this
additional information on the $Q\bar{q}$ system will enable a less ambiguous
model of this system to be made. At present, if through an operator
$O(r)$, a transition between
two two-quark states is calculated as
$\langle \psi_1(r)|O(r)|\psi_2(r)\rangle$, then the
$\psi_i$ are first calculated as a byproduct of some differential
equation containing some interquark potential - say $V(r)=a/r+br$.
However, both the form of $V(r)$ and the equation into which
it is inserted are not unique. The latter could range from a non-relativistic
Schroedinger equation, if the quarks are heavy enough, through a series
of equations that incorporate relativistic effects to varying degrees.

\section{Heavy-light mesons $(Q\bar{q})$}\label{hl}
A study of heavy-light mesons is not just for the academic interest
discussed above, since they are realised in nature. The best examples
are the $B(5.28 \ {\rm GeV})$ and $B_s(5.37 \ {\rm GeV})$ mesons, which have quark
structures $(\bar{b}u)$ and$(\bar{b}s)$. Since the $b, \ s $ and
$u$-quarks have masses of about $4.2, \ 0.1$ and 0.001 GeV, we see that
the $B$ and $B_s$ are indeed heavy-light mesons and can be thought of as
the "Hydrogen atom" for quark systems. Currently the $B$-mesons are of
particular interest, since they are
expected to lead to a better understanding of CP violation
and for this reason are  being generated at so-called $B$-factories\cite{bf}.

More details of  the following sections can be found in
Ref.\cite{gkpm}.

\subsection{Energies of Heavy-light mesons}
\begin{figure}
\centering
\epsfxsize=17pc
\epsfbox{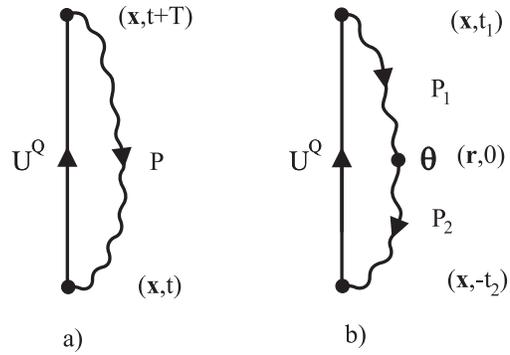}
\caption{a) A two-point correlation function, b) A three-point
correlation function. \label{fig.C2C3}}
\end{figure}

The basic quantity for evaluating the energies of heavy-light mesons is
the 2-point correlation function -- see Fig.~\ref{fig.C2C3} a).
\be
C(2,T)=\langle U^Q({\bf x},t,T)P({\bf x},t+T,t)\rangle,
\ee
where $U^Q({\bf x},t,T)$ is the heavy(infinite mass)-quark propagator
and $P({\bf x},t+T,t)$ the light-quark propagator. The $\langle ...\rangle$
means that $C(2)$ has been averaged over the whole lattice.
Since the $C(2)$ decay as $\exp(-E_0T)$, where $E_0$ is the energy of
the ground state, we get
\be
\label{E_0}
E_0=- \ln \ [\frac{\langle C(2,T)\rangle}{\langle
C(2,T-1)\rangle}] \  \  \ {\rm as}
\ T\rightarrow \infty.
\ee
These energies were calculated in Ref.\cite{MP} for the states

$L_J=S_{1/2}, \ P_{1/2}, \ P_{3/2}, \ D_{3/2}, \ D_{5/2}, \ F$

\subsection{Heavy-Light radial correlations}
Here the basic quantity is  3-point correlation function
 -- see Fig.~\ref{fig.C2C3} b).
\be
C(3,-t_2, \ t_1, \ {\bf r})=\langle U^Q({\bf x},-t_2, \ t_1)
P_1({\bf x},t_1;{\bf r}, 0) \Theta ({\bf r}) P_2({\bf r},0;{\bf
x},-t_2)\rangle,
\ee
where the $P_{1,2}$ are the light anti-quark propagators that go from
the $Q$ at time $t_1$ to the point ${\bf r}$ at $t=0$ and then return to
$Q$ at time $-t_2$. The probe $\Theta ({\bf r})$ at ${\bf r}$ is here
considered to have two forms
i) $\Theta=\gamma_4$ for measuring the charge distribution of the $\bar{q}$ and
ii) $\Theta=1$ for measuring the matter  distribution.

Knowing the $C(3)$ then the radial distributions are given by
\be
\label{FC}
F[C(\Theta, T,R)]=\frac{\langle C(3,\Theta,T,R)\rangle}{\langle
C(2,T)\rangle}.
\ee

\subsection{ Lattice Parameters and Refinements}
The essential lattice parameters are:

1) Lattice size $16^3\times 24$

2) Quark-gluon coupling $\beta=5.7$ giving
a lattice spacing $\approx 0.17$fm

3) Light quark mass $\approx m_s$ -- see Ref.~\cite{shan} as evidence
for this.

4) Quenched Approximation
 i.e. quark-antiquark pairs are not included.

\noindent Eqs.~\ref{E_0} and \ref{FC} for the ground state energy($E_0$) and
 correlations $F[C(\Theta, T,R)]$
are based on a single state constructed from the basic lattice.
However, this can be extended by modifying this original lattice to give
other states ($i,j,k...$). In this way the 2- and 3-point correlation
functions can become matrices $C_{i,j}(2), \ \ C_{i,j}(3,{\bf r})$.
Their diagonalisation -- by reducing the contamination of higher states on the
ground state -- then gives improved estimates of $E_0$ and
$F[C(\Theta, T,R)]$.
The mechanism for generating these addition states is fuzzing.
This replaces a  link on the lattice by a combination of neighbouring
links  to give a fuzzed link -- with projection to SU3 implied -- as  

[A fuzzed link] = $f_p\cdot $[Straight link]+ [Sum of 4 spatial
U-bends].

This cycle can be repeat many times.
Here, in addition to the basic state(L),
 two new states F1 and F2 are constructed using 2 and 6 cycles with $f_p=2.5$.
respectively.
\section{Results}
The lattice data  $\langle C_{i,j}(2,T)\rangle$ and
$\langle C_{i,j}(3,T,{\bf r})\rangle$ can be analysed in several ways.
Here two are presented: i) Visual and  ii) Fitting with exponentials.
\subsection{Visual}
\begin{figure}
\centering
\begin{minipage}[c]{0.5\textwidth}
\centering
\epsfxsize=13.5pc
\epsfbox{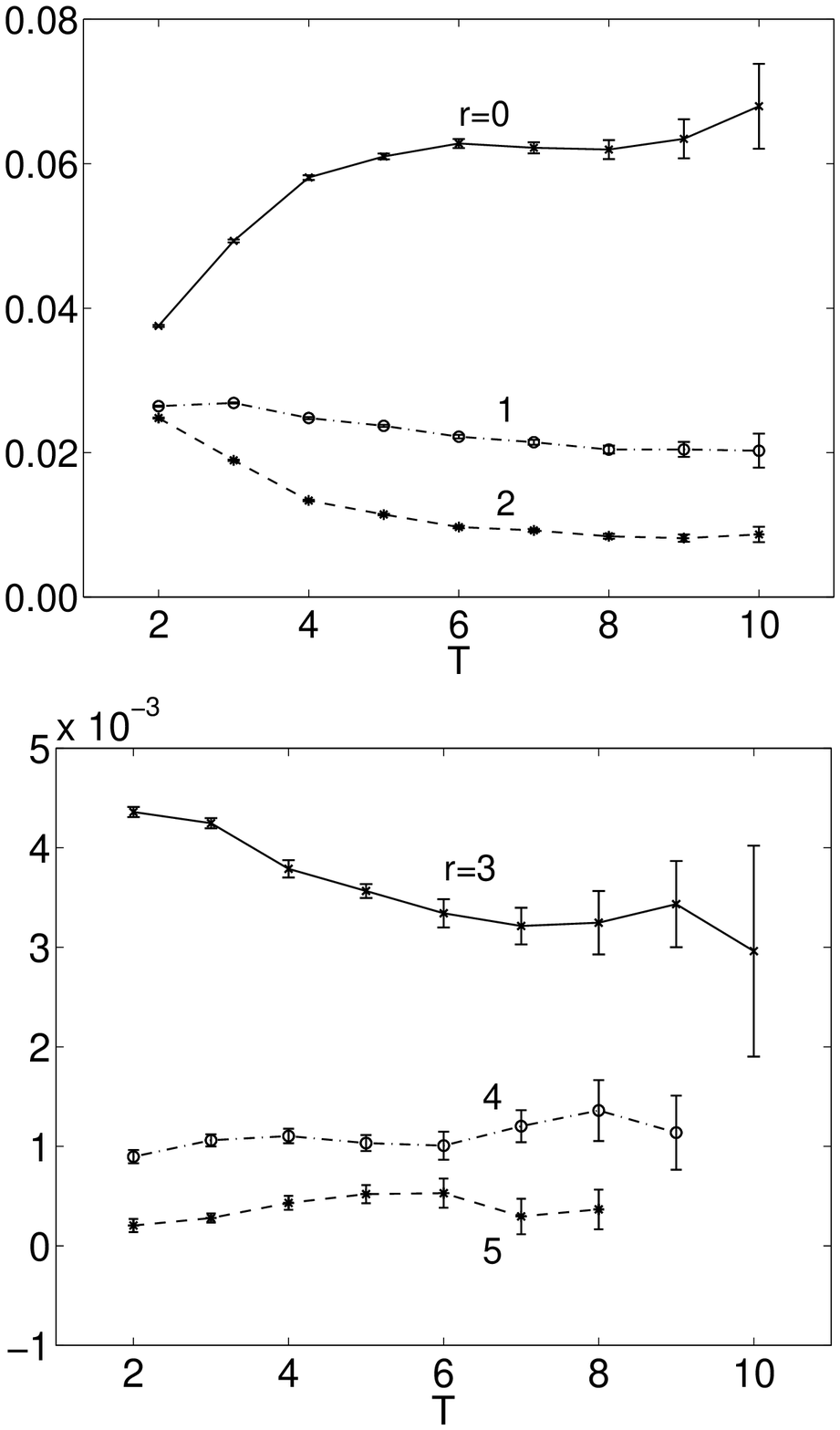}
\caption{The ratio $\langle C(3,r)\rangle/\langle C(2)\rangle$
for \newline $r=0,\ldots ,5$. \label{fig.r}}
\end{minipage}%
\begin{minipage}[c]{0.5\textwidth}
\centering
\epsfxsize=13.5pc
\epsfbox{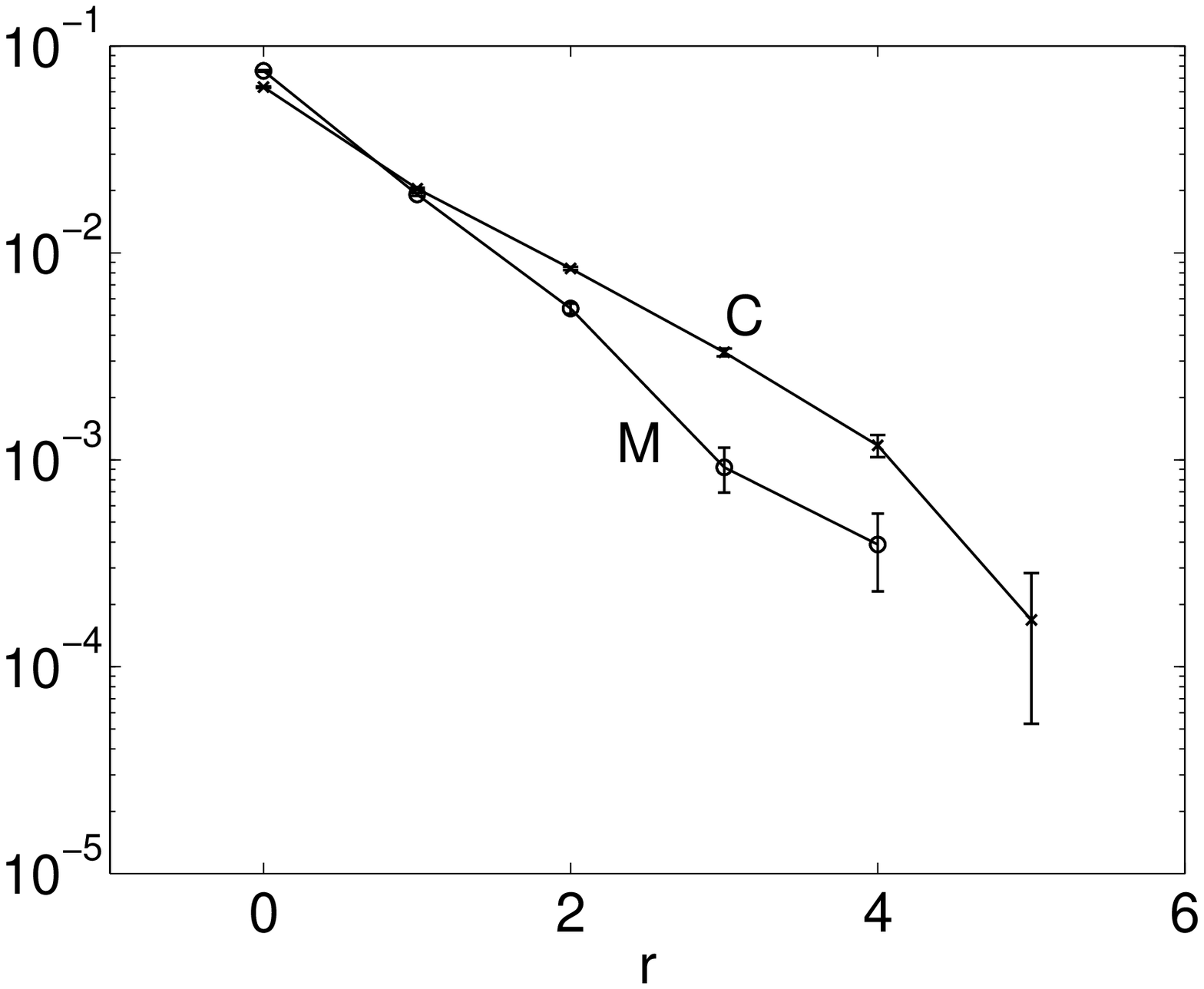}
\caption{The radial distribution of the ground state
charge(C) and matter(M) densities. \label{wr}}
\end{minipage}
\end{figure}

 Fig.~\ref{fig.r} shows from Eq.~\ref{FC} the ratio
$\langle C(3,R)\rangle/\langle C(2) \rangle$ for values of $r$ upto 5
lattice spacings i.e. about 0.8fm.  As $T\rightarrow \infty$, these clearly
show plateaux, which give directly the desired charge or matter density.
\subsection{Fitting with exponentials}
A more precise method is to first fit the $\langle C_{i,j}(2,T)\rangle$
by the approximate expression
\be
\tilde{C}_{ij}(2, T)=\sum_{\alpha =1}^{3}v_i^{\alpha}
\exp(-E_{\alpha}T)v_j^{\alpha},
\ee
where the $i,j$ refer to the states L,F1,F2. This results in the energy
eigenvalues $E_{\alpha}$ and their eigenvectors $v_i^{\alpha}$. These
$E_{\alpha}$ are the energies quoted in Ref.~\cite{MP}. It is found that
only data with  $T\ge 4$ gives $\chi^2/dof\approx 1$. This means that
54 data points are fitted with 12 parameters.

Given the $E_{\alpha}$ and $v_i^{\alpha}$ the $C_{ij}(3,T,R)$ are then
fitted by the approximate expression
\be
\tilde{C}_{ij}(3,T,r)= \sum_{\alpha =1}^{3}\sum_{\beta =1}^{3}
v_i^{\alpha} \exp[-E_{\alpha}t_1]x^{\alpha\beta}(r)
\exp[-E_{\beta}t_2]v_j^{\beta},
\ee
where the $x^{\alpha\beta}(r)$ are varied giving directly the desired
charge or matter density.
This needs $t_1+t_2 \ge 8$ to get $\chi^2/dof\approx 1$
when fitting 18 data points with 3 parameters.
\subsection{Analytic forms for the charge and  matter densities}
For the above charge $(i$=$c)$ and  matter $(i=m)$ densities, it is of
interest to express $x_i(r)=w_i^2(r)$ in the form
$w_i(r)=A_i\exp(-r/r_i$), where the parameters $r_i$ acd $A_i$ are given
for a global fit in Table~\ref{t1}.
\begin{table}[t]
\label{t1}
\caption{The parametrization of the charge and matter
densities as $w^2_m(r)$, where
 $w_m(r)=A_m\exp[-r/r_m]$. }
\begin{center}
\footnotesize
\begin{tabular}{|l|c|c|c|}
\hline
&$A_i$&$r_i$ (fm)&RMS Radius (fm)\\ \hline
Charge Density &0.26(1)&0.32(2)&0.55(3)\\
Matter Density &0.29(1)&0.24(2)&0.42(3)\\
\hline
\end{tabular}
\end{center}
\end{table}
There, as in Fig.~\ref{wr}, it is clearly seen that the
charge density has a significantly
longer range than that of the matter density.

The fourier transforms of these densities lead to  vector$(i$=$v=c)$ and
scalar$(i=s=m)$ form factors that are usually expressed as
$[F_{v,s}(q^2)] \propto (q^2+M_{v,s}^2)^{-1}.$
These are appropriate for the long-range part of $w_i^2(r)$ and lead to
$M_v$=0.9(1) and $M_s=1.3(1)$GeV and should be compared with the
the direct calculation of the $s\bar{s}$ vector and scalar mesons in
Refs.~\cite{mm} and \cite{shan}. The latter only involve 2-point
correlation functions and so are more precise leading to $M_v$=0.944(2)
and $M_s=1.61(6)$GeV.
\subsection{Sum-rules}
The above measures the densities at a few definite values of $r$.
However, it is of interest to consider the sum-rules that sum over all
values of $r$ i.e.
\be
F^{{\rm sum}}[C(3)]=\frac{\langle\sum_{{\bf r}}C(3,{\bf r})\rangle}
{\langle C(2)\rangle}.
\ee
For the charge and matter densities this yields $I_c$=1.30(5) and $I_m$=0.4(1)
respectively. On the other hand, integrating the above analytic
expressions for $w_i(r)$ gives 1.5(1) and 0.7(1). These estimates
are less reliable than the direct lattice sum, since much of the
contributions are from small values of $r$ where lattice artifacts
enter.

By charge conservation we should get $I_c$=1 in the continuum limit --
the quark charge having been defined as unity.
However, it is known that renormalisation effects enter due to the
finite lattice spacing. In Ref.~\cite{bo}, for lattice parameters
different to those used here, an overall renormalisation factor ($F^V$) of
$\approx 0.8$ is found for the vector(charge) vertex. Similar reductions
are found in Ref.~\cite{Div} for the axial vector operator. The inverse
of our value of $I_c$=1.30(5) could then be interpreted as a
non-perturbative estimate of $F^V$ as 0.77(3). In any case the
conclusion is that our $I_c$=1.30(5) is consistent with charge
conservation for the continuum.
\subsection{Dirac amplitude interpretation}
In a relativistic quantum mechanical picture the above charge and matter
distributions can be interpreted as
$|g|^2+|f|^2$ and $|g|^2-|f|^2$, where $g,f$ are the upper and lower
components in the solution of the Dirac equation. Possibly this will now serve
as a way to remove some of the ambiguities refered to at the end of
Section~\ref{intro} by ensuring that the solutions of the Dirac equation
agree not only with the lattice or empirical energies but also the forms
of $|g(r)|^2$ and $|f(r)|^2$.
\section*{Conclusions and Future}
\begin{itemize}
\item  The $S_{1/2}$-wave charge and matter densities can be measured quite
reliably out to $\approx 0.7$fm.
\item Off-axis measurements can improve these results. In particular,
it is interesting to see whether or not the charge density for $r=5$
is indeed lower than the simple exponential drop off in Fig.~\ref{wr}.
This would indicate that the confining potential is playing a role.
Hopefully, the better statistics at the off-axis point $(3,4)$ would
clarify this point.
\item The $P_{1/2}, \ P_{3/2}, \ D_{3/2}, \
D_{5/2},...$ densities are now being measured.
\item Understand the densities phenomenologically using the Dirac equation.
\item Measure correlations in  the $(Q^2\bar{q}^2)$ system
and check their form against the above $Q\bar{q}$-correlations.
\item Use larger $\beta$ and lattices to get nearer the continuum limit.
\item Replace the present quenched lattices by unquenched ones.
\item Consider other operators, e.g. Pseudo-vector
$(\gamma_{\mu}\gamma_5)$ for $B^*B\pi$ coupling.
\end{itemize}
\section*{Acknowledgments}
The authors wish to thank the Center for Scientific Computing in Espoo,
Finland for making available resources without which this project could
not have been carried out.

\begin{thebibliography}{99}
\bibitem{glo}W. Gl\"{o}ckle and H. Kamada \Journal{\em Phys.Rev.Lett.}
{71}{971}{1993}
\bibitem{carl} J. Carlson and R. Schiavilla ,
\Journal{ {\em Rev.Mod.Phys}} {70} {743}{1998}

\bibitem{mos}S.A. Moszkowski and B.L. Scott,
 \Journal{ {\em Ann. of Phys}} {36}{2109}{1987}.

\bibitem{PG} P. Pennanen and A.M. Green ,
\Journal{\em Phys. Rev.} {57} {3384} {1998}.

\bibitem{GKP} A.M. Green, J. Koponen and P. Pennanen,
\Journal{\PRD} {61} {014014} {2000}.

\bibitem{PGM} P. Pennanen, A.M. Green and C. Michael 
\Journal{\PRD} {59} {014504} {1999}.

\bibitem{bf} H. Quinn and M. Witherell,  \Journal {\em Sc.Am}{279}{50}
{1998}
\bibitem{gkpm}A.M. Green, J. Koponen, P. Pennanen and C. Michael,
"The Charge and Matter radial distributions of Heavy-Light mesons
calculated on a lattice", hep-lat/0105027

\bibitem{MP}
C.~Michael and J.~Peisa, \Journal{\PRD}{D58}{034506} {1998}
{\tt hep-lat/9802015}
\bibitem{mm} UKQCD Collaboration, C.~McNeile and C.~Michael,
{\tt hep-lat/0010019}
\bibitem{shan} UKQCD Collaboration, H. Shanahan et al.,  \Journal{\PRD}
{D55} {1548}{1997}
\bibitem{bo} K.C.~Bowler, L.~Del Debbio, J.M.~Flynn, G.N.~Lacagnina,
V.I.~Lesk, C.M.~Maynard and D.G.~Richards, {\tt hep-lat/0007020}
\bibitem{Div} G.M.~de Divitiis, L.~Del Debbio, M.~Di Pierro,
J.M.~Flynn, C.~Michael and J.~Peisa, JHEP 9810:010,1998
{\tt hep-lat/9807032}
\
\end{thebibliography}
\end{document}